\documentclass[conference,letterpaper]{IEEEtran}

\usepackage{graphicx}
\usepackage{amsmath,amssymb}
\usepackage{booktabs}
\usepackage{multirow}
\usepackage{makecell}
\usepackage{float} 
\usepackage{pifont}
\usepackage{xcolor}
\usepackage[numbers,sort&compress]{natbib}

\newcommand{\cmark}{\ding{51}}
\newcommand{\xmark}{\ding{55}}
\providecommand{\etal}{\emph{et al.}}
\usepackage[unicode,hidelinks]{hyperref}

\title{CS-LDM: Clinically Guided Structure-Aware
Latent Diffusion for Knee Radiographic
Forecasting}

\author{
\IEEEauthorblockN{Shaotian Li\IEEEauthorrefmark{1}, Chenhui Wang\IEEEauthorrefmark{1}, Wenhua Wu\IEEEauthorrefmark{2}, Yanqiu Wu\IEEEauthorrefmark{1}, Xiaohan Yu\IEEEauthorrefmark{1}}
\IEEEauthorblockA{\IEEEauthorrefmark{1}Macquarie University\\
\texttt{leslie.li@mq.edu.au, yanqiu.wu@mq.edu.au, xiaohan.yu@mq.edu.au}}
\IEEEauthorblockA{\IEEEauthorrefmark{2}University of Sydney\\
\texttt{wenhua.wu@sydney.edu.au}}
}

\begin{document}
\maketitle

\begin{abstract}
Knee osteoarthritis (KOA) is a leading cause of physical disability and currently has no curative clinical treatment. Early diagnosis and accurate prognosis are therefore vital for managing its progression. Existing methods mainly predict future severity grades and provide limited visual explanation of structural change. Recent generation-guided approaches synthesize follow-up radiographs for progression assessment, but generated images are often judged by predictive accuracy or image realism, so they may satisfy a grading model while remaining morphologically unreliable. We therefore propose CS-LDM, a clinically guided, structure-aware latent diffusion framework that forecasts patient-specific future knee radiographs from the current X-ray and clinically available current KL grade, using LoRA adaptation, current-grade conditioning, and frozen-teacher diagnostic distillation. Experiments on the public Osteoarthritis Initiative (OAI) dataset show that CS-LDM improves future severity-grade consistency while maintaining competitive image fidelity. Further ordinal, OARSI, ROI-level, and stable/progressed analyses show that radiographic forecasting should be evaluated by balancing diagnostic consistency, anatomical plausibility, and structural preservation.
\end{abstract}

\section{Introduction}
\label{sec:intro}

Knee osteoarthritis (OA) is a leading cause of chronic pain, long-term disability, and structural joint degeneration worldwide. Accurate assessment of disease progression is important for early intervention, clinical trial stratification, and patient monitoring. In routine radiographic assessment, disease severity is commonly summarized using the Kellgren--Lawrence (KL) grading system \cite{kellgren1957radiological}, which reflects ordered structural changes such as joint-space narrowing, osteophyte formation, and subchondral sclerosis. Mirroring this clinical convention, many learning-based OA studies formulate radiographic assessment as static KL classification~\cite{antony2016quantifying, tiulpin2018automatic, tiulpin2020automatic} or multi-modal scalar risk prediction~\cite{tiulpin2019multimodal, hirvasniemi2023knee, bayramoglu2024deep, panfilov2025end}. Although these models offer useful quantitative predictions, they compress complex, high-dimensional structural change into a single label or risk score. As a result, they provide limited insight into where and how an individual patient's joint morphology may change over time. This gap between numerical risk estimation and patient-specific visual interpretation raises a broader question: \textit{Can radiographic OA progression be forecasted in a way that is both quantitatively informative and visually interpretable?}


An emerging line of research has begun to address this limitation through generation-guided OA modeling. For instance, IC-RDN~\cite{wu2024identity} synthesizes future knee radiographs to assist downstream progression grading, whereas Butler \etal~\cite{butler2025risk} leverages predictive diffusion architectures for visual risk estimation. These approaches provide a more interpretable intermediate representation than black-box scalar prognosis, since the generated follow-up image can be visually inspected. However, they also introduce an evaluation challenge that high downstream classification accuracy does not necessarily imply anatomical validity. When generated radiographs are judged mainly by learned grading networks, the output may become aligned with an evaluator's decision boundary while still containing shortcut artifacts, unrealistic texture, or distorted joint morphology. Therefore, longitudinal radiographic forecasting requires not only diagnostic consistency, but also evidence that synthesized changes occur in clinically meaningful anatomical regions.


This challenge is further compounded by broader medical image-to-image (I2I), restoration, and longitudinal synthesis frameworks. Sequential diffusion architectures model large-scale disease trajectories~\cite{yoon2023sadm, dao2024conditional, cao2024medical}, while general generative networks mainly address style transfer or restoration~\cite{isola2017image, zhu2017unpaired, muller2023multimodal, xia2024diffi2i, zou2025cyclediff}. Specialized medical I2I diffusion models~\cite{jiang2025fast, xie2025meddiff, zhang2025geodesicdiffusionmodelsmedical, arslan2024selfconsistent} are designed for static modality translation or enhancement rather than patient-specific temporal forecasting. Restoration-style diffusion backbones recover a high-quality image from a degraded observation, where the input and target share the same anatomical state~\cite{lin2025harnessing}. In contrast, 12-month knee radiographic forecasting requires future-state synthesis. Specifically, the model must preserve baseline skeletal morphology while allowing exceptionally subtle, localized OA-related changes such as joint-space narrowing and osteophyte development. Without explicit clinical and structural constraints, generic generation models may suppress progression cues or introduce global distortions that disrupt patient-specific anatomy.


To address this methodological tension, we reframe the relationship between generative models and clinical classifiers. External classifiers should not function as rigid optimization targets that may encourage classifier-specific image changes~\cite{dhariwal2021diffusion}. Instead, clinical diagnostic knowledge can be used as a soft semantic regularizer that guides the generator toward future KL-relevant radiographic cues while preserving patient-specific structure. This design aims to avoid treating classifier confidence as the sole objective, and instead balances two requirements: the generated follow-up should remain anatomically plausible with respect to the baseline image, and it should contain diagnostic evidence consistent with the real 12-month follow-up state.


Based on this insight, we propose CS-LDM, a clinically guided latent forecasting framework for patient-specific 12-month knee radiographic forecasting. Given a baseline knee radiograph, CS-LDM synthesizes a follow-up X-ray that remains anchored to the patient's current morphology. Structurally, CS-LDM repurposes a restoration-style latent diffusion prior~\cite{rombach2022high, ho2020denoising} for future-state radiographic synthesis, adapting the backbone with low-rank adaptation (LoRA)~\cite{hu2022lora}. The generation process is conditioned on the clinically observed baseline KL grade, which provides an available severity prior without requiring oracle access to the future grade. Semantically, we introduce frozen-teacher soft diagnostic distillation, where a KL grading network trained exclusively on real follow-up radiographs provides soft diagnostic supervision~\cite{hinton2015distilling}. This mechanism aligns prediction distributions between real and generated 12-month images, encouraging future KL-relevant radiographic cues without forcing the image toward a discrete class prototype.


We evaluate CS-LDM on the longitudinal Osteoarthritis Initiative (OAI) cohort~\cite{nevitt2006osteoarthritis}. Because generated radiographs can become classifier-legible without being anatomically faithful, we separate automated diagnostic consistency from anatomical-change fidelity. Diagnostic consistency is assessed using real-image-trained KL graders across multiple architectures and ordinal KL metrics. Anatomical fidelity is assessed using OARSI joint-space narrowing and osteophyte agreement, zoomed ROI visualizations, and V00--V12 difference-map analysis. We further report stable/progressor stress testing with a no-change baseline, treating binary progression detection as secondary due to the small number of true progressors. CS-LDM improves future KL-grade consistency while preserving patient-specific morphology and localizing changes to clinically meaningful regions. We therefore interpret improved automated KL accuracy as diagnostic consistency, not clinical superiority over real follow-up radiographs.

Our main contributions are summarized as follows: 
\begin{itemize} 
    \item We formulate knee OA prognosis as patient-specific longitudinal
    radiographic forecasting, producing an inspectable 12-month image rather than
    only a scalar grade.

    \item We propose CS-LDM, a clinically guided latent forecasting framework that repurposes a restoration-style diffusion prior for future-state radiographic synthesis, using only baseline-available clinical information and soft diagnostic distillation without future-label leakage.

    \item We introduce an evaluation protocol that separates diagnostic consistency
from anatomical-change fidelity using cross-architecture KL graders, OARSI
feature agreement, ROI difference maps, and stable/progressed stress testing.
\end{itemize}





\section{Related Work}
\label{sec:related}

\subsection{Knee OA Evaluation and Generation-Guided Forecasting}

Conventional deep learning workflows for knee osteoarthritis (OA) on the Osteoarthritis Initiative (OAI) cohort predominantly focus on automated diagnostic grading or scalar risk estimation. Following clinical conventions standardized by the Kellgren--Lawrence (KL) system~\cite{kellgren1957radiological}, early convolutional architectures formulate radiographic assessment as static classification, mapping current joint morphology to discrete severity grades~\cite{antony2016quantifying, tiulpin2018automatic, tiulpin2020automatic}. To capture temporal dynamics, later multimodal progression frameworks incorporate longitudinal clinical variables alongside imaging to predict scalar progression probabilities or multi-year risk scores~\cite{tiulpin2019multimodal, hirvasniemi2023knee, bayramoglu2024deep, panfilov2025end}. While useful for patient stratification, these outputs compress complex structural evolution into a single index and provide limited visual explanation of how an individual patient's joint may change.

To improve interpretability, recent work has explored future radiograph generation for OA prognosis. IC-RDN~\cite{wu2024identity} uses an identity-prior diffusion framework to forecast radiographs for downstream progression grading, while Butler \etal~\cite{butler2025risk} combines predictive diffusion with visual risk estimation and landmark-based biomarkers. These methods establish future-image synthesis as a useful intermediate representation between raw imaging and scalar prognosis. However, clinical validity is assessed primarily through downstream classifier performance or qualitative visual comparison. This leaves open whether improved classification reflects anatomically meaningful change, such as joint-space narrowing and osteophyte development, or alignment with the evaluator's learned decision boundary. In contrast, our work separates automated KL consistency from anatomical-change fidelity using OARSI sub-feature agreement~\cite{tiulpin2020automatic} and ROI-level V00--V12 difference-map analysis.

\subsection{Longitudinal Medical Image Generation}

Beyond musculoskeletal imaging, generative models have been used to represent disease trajectories and anatomical change over time. Longitudinal architectures have been proposed for tasks such as brain atrophy modeling, progressive neurological change, and long-term organ aging simulation~\cite{yoon2023sadm, dao2024conditional, cao2024medical}. For example, SADM~\cite{yoon2023sadm} models temporal dependencies with sequence-aware diffusion conditioning, while related approaches synthesize continuous medical image trajectories from a single scan using visit-conditioned generation~\cite{dao2024conditional, cao2024medical}. These frameworks are commonly designed for multi-timepoint 3D volumetric sequences, such as structural MRI, where anatomical trajectories can be modeled over longer horizons. Knee OA forecasting presents a different setting. Over a 12-month window, radiographic progression is often subtle, localized, and constrained by the patient's baseline skeletal structure. Relevant changes are concentrated around the joint space and osteophyte-prone margins, while most patient-specific morphology should remain stable. As a result, unconstrained longitudinal generators may either suppress subtle progression cues or drift from the baseline anatomy. CS-LDM addresses this setting by adapting a restoration-style latent prior for structure-preserving future-state synthesis.

\subsection{Medical Image-to-Image Synthesis Baselines}

Our framework is also related to medical image-to-image (I2I) translation.  Classical paired conditional GANs such as Pix2Pix~\cite{isola2017image} and unpaired mapping models such as CycleGAN~\cite{zhu2017unpaired} established the foundations for cross-domain translation, although adversarial training may introduce hallucinated structures or fail to preserve patient-specific anatomy. More recent diffusion-based translators, including DiffI2I~\cite{xia2024diffi2i} and CycleDiff~\cite{zou2025cyclediff}, improve high-fidelity image translation and restoration. Within the medical domain, specialized I2I diffusion models are widely used for cross-modality conversion, image enhancement, and synthetic data augmentation~\cite{muller2023multimodal, jiang2025fast, xie2025meddiff, zhang2025geodesicdiffusionmodelsmedical, arslan2024selfconsistent, sagers2023augmenting, saragih2024using}. Specifically, Fast-DDPM~\cite{jiang2025fast} optimizes diffusion sampling speed for medical translation, GDM-VE~\cite{zhang2025geodesicdiffusionmodelsmedical} leverages geodesic diffusion schedules for conditional enhancement, and SelfRDB~\cite{arslan2024selfconsistent} formulates translation as a distribution bridge.

Although these models are meaningful baselines for learning image mapping distributions, they are primarily designed for stationary translation, enhancement, or restoration. Longitudinal knee forecasting is different because the output should remain anatomically tied to the baseline image while reflecting plausible future OA-related change. We therefore use these models as comparative baselines, but distinguish them from methods specifically designed for paired current-to-follow-up forecasting. Generic text-driven editing methods such as FBSDiff~\cite{gao2024fbsdiff} and mask-conditioned lesion synthesis methods such as MedDiff-FT~\cite{xie2025meddiff} are discussed as related generative works rather than direct baselines, since they do not learn patient-specific V00-to-V12 radiographic forecasting.

\subsection{Classifier-Guided Generation and Distillation}



External networks are commonly used to control generative models through classifier guidance, which steers diffusion sampling toward target attributes during inference~\cite{dhariwal2021diffusion}, or through knowledge distillation, where a student model learns from softened teacher outputs~\cite{hinton2015distilling}. In medical image generation, however, directly optimizing toward classifier responses can make generated images more classifiable without ensuring anatomical validity. This is especially problematic for radiographic forecasting, where small image changes may strongly influence grading predictions while not corresponding to plausible disease progression.

CS-LDM differs from hard classifier guidance and conventional parameter-compression distillation. We do not use a classifier to guide each sampling step at inference. Instead, a frozen KL-grade teacher is used only during training as a soft diagnostic regularizer. The teacher distribution computed from the real follow-up image provides a soft target for the generated follow-up, encouraging radiographic evidence without forcing the image into a discrete class prototype. This design integrates clinical semantics while preserving the structure-aware forecasting objective.

\section{Methodology}
\label{sec:method}

\begin{figure*}[t]
    \centering
    \includegraphics[width=\linewidth]{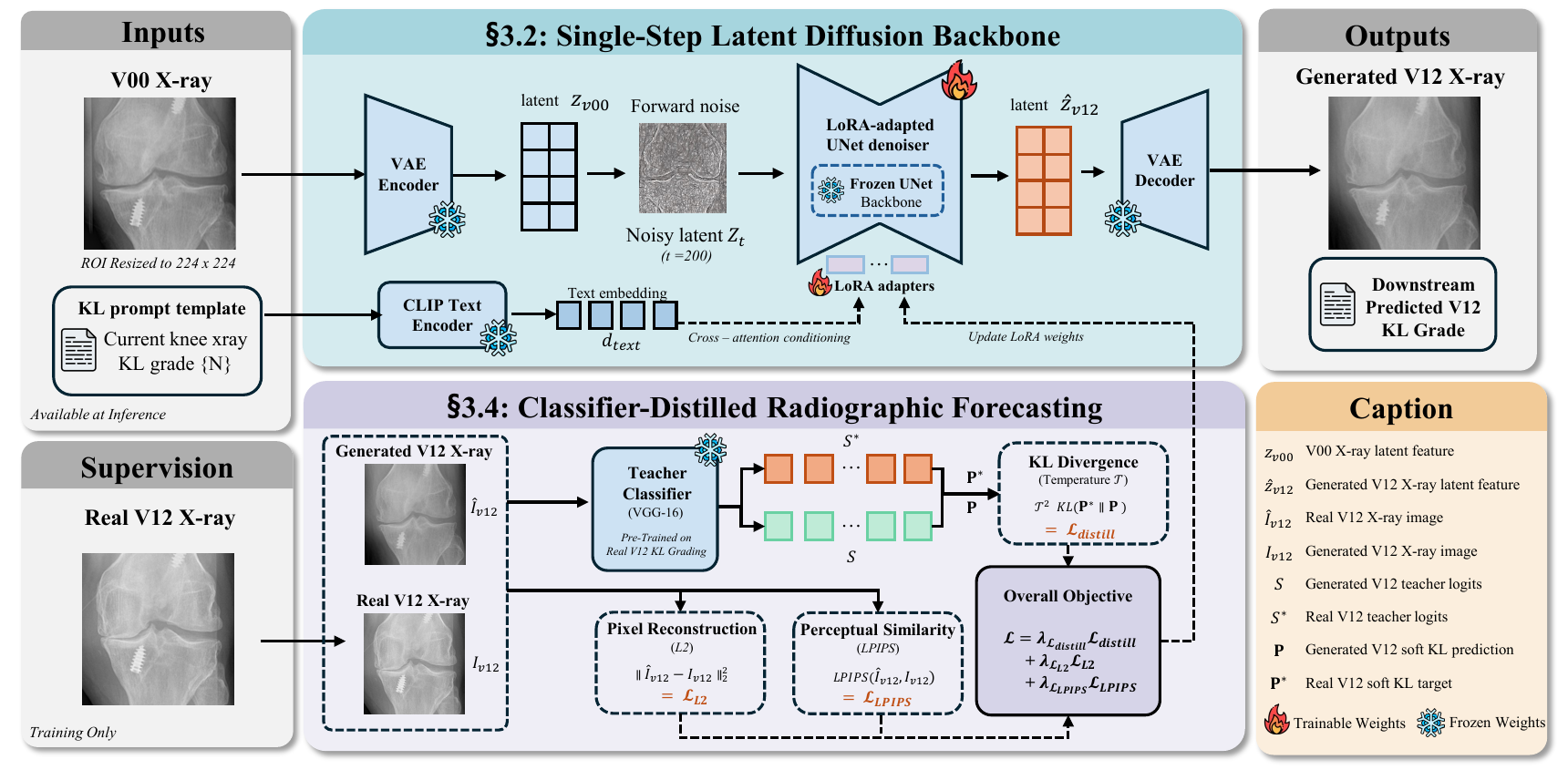}
    \caption{Overview of CS-LDM, a clinically guided latent forecasting framework for synthesizing patient-specific 12-month follow-up knee radiographs from current X-rays.}
    \label{fig:method_pipeline}
    \vspace{-1em}
\end{figure*}

\subsection{Problem Formulation}
\label{sec:problem}

Let $I^{i}_{v00} \in \mathbb{R}^{C \times H \times W}$ denote the current knee
radiograph of joint $i$, and let $I^{i}_{v12}$ denote the corresponding
12-month follow-up radiograph. Each joint has Kellgren--Lawrence (KL) grades
$y^{i}_{v00}, y^{i}_{v12} \in \{0,1,2,3,4\}$ at the two visits. We refer to the
V00 visit as the current, or baseline, visit; throughout the experiments, the
model receives only the V00 radiograph and V00 KL grade at inference.

Our goal is to learn a patient-specific longitudinal generator
\begin{equation}
    G_{\theta}: I^{i}_{v00} \rightarrow \hat{I}^{i}_{v12},
\end{equation}
where $\hat{I}^{i}_{v12}$ is a synthesized future radiograph. The generated
image should satisfy two requirements: it should preserve the patient-specific
morphology visible in the current scan, and it should express KL-relevant
radiographic cues associated with the 12-month follow-up state.

We define this task as longitudinal radiographic forecasting. Unlike direct
prognosis models that output only a future label, our method predicts an
intermediate future image. This allows both visual inspection and downstream
clinical evaluation. We evaluate generated follow-ups from two complementary
perspectives: automated diagnostic consistency, measured by real-image-trained
KL graders, and anatomical-change fidelity, assessed through image similarity,
feature-level agreement, and local structural analyses against the true
follow-up $I^{i}_{v12}$.

\subsection{Latent Forecasting Backbone}
\label{sec:backbone}

We use a latent diffusion backbone based on Stable Diffusion 2.1~\cite{rombach2022high}. The model consists of a variational autoencoder (VAE)~\cite{kingma2013auto} with encoder $\mathcal{E}$ and decoder $\mathcal{D}$, a UNet denoiser $\epsilon_{\phi}$, and a CLIP text encoder $\tau$. To adapt the pretrained model for baseline-to-follow-up radiographic forecasting, we insert Low-Rank Adaptation (LoRA) modules into the UNet and optimize only the LoRA parameters $\theta$. The pretrained VAE, UNet, and text encoder weights are kept frozen.

Given a current radiograph $I_{v00}$, we first encode it into the latent space:
\begin{equation}
    z_{v00} = \mathcal{E}(I_{v00}).
\end{equation}
We then perturb the latent representation to a fixed timestep $t$ using the standard diffusion forward noising process:
\begin{equation}
    z_t =
    \sqrt{\bar{\alpha}_t} z_{v00}
    +
    \sqrt{1-\bar{\alpha}_t}\epsilon,
    \qquad
    \epsilon \sim \mathcal{N}(0,I).
\end{equation}
The LoRA-adapted UNet predicts the future follow-up latent in a single denoising
step:
\begin{equation}
    \hat{z}_{v12} = f_{\theta}(z_t, t, c),
\end{equation}
where $c$ is the text-conditioning embedding and $f_{\theta}$ denotes the
single-step denoising mapping induced by the frozen UNet and trainable LoRA
adapters. The synthesized future radiograph is obtained by decoding:
\begin{equation}
    \hat{I}_{v12} = \mathcal{D}(\hat{z}_{v12}).
\end{equation}

The timestep $t$ design restricts the model to a structure-preserving
transformation of the current joint structure rather than unconstrained synthesis.
This is important for medical forecasting, where hallucinated morphological changes
could lead to clinically misleading predictions.

\subsection{Current KL-Grade Conditioning}
\label{sec:conditioning}

To provide a weak clinical prior while keeping the model usable at inference
time, we condition the generator on the known current KL grade. For each joint
$i$, we construct the prompt
\begin{equation}
    p^{i} =
    \text{``knee xray kl grade } y_{v00}^{i}\text{''},
\end{equation}
and encode it using the frozen text encoder:
\begin{equation}
    c^{i} = \tau(p^{i}).
\end{equation}
The resulting embedding is injected into the UNet through cross-attention. Since
$y_{v12}^{i}$ is unavailable during deployment, we condition only on
$y_{v00}^{i}$, making the prompt a clinically available prior rather than an
oracle condition.

\subsection{Classifier-Distilled Radiographic Forecasting} \label{sec:distillation} Pixel-level and perceptual reconstruction losses encourage visual similarity, but they do not guarantee that the generated image preserves diagnostic features needed for KL grading. We therefore introduce a classifier-distillation objective using a frozen KL-grade teacher. We first train a VGG-16 classifier $T_{\psi}$ on real 12-month radiographs for 5-class KL grading. During generator training, $T_{\psi}$ is frozen and used only as a fixed diagnostic teacher. For a synthesized image $\hat{I}_{v12}$ and its real follow-up target $I_{v12}$, we compute the teacher logits \begin{equation} \mathbf{s} = T_{\psi}(\hat{I}_{v12}), \qquad \mathbf{s}^{\star} = T_{\psi}(I_{v12}), \end{equation} where $\mathbf{s}^{\star}$ is detached and serves as a  soft diagnostic target. We then define the softened KL-grade distributions as \begin{equation} \mathbf{p} = \sigma(\mathbf{s}/\mathcal{T}), \qquad \mathbf{p}^{\star} = \sigma(\mathbf{s}^{\star}/\mathcal{T}), \end{equation} where $\sigma(\cdot)$ is the softmax function and $\mathcal{T}$ is the distillation temperature. The classifier-distillation loss is \begin{equation} \label{eq:distill} \mathcal{L}_{\mathrm{distill}} = \mathcal{T}^{2} \mathrm{KL} \left( \mathbf{p}^{\star} \; \middle\| \; \mathbf{p} \right). \end{equation} The factor $\mathcal{T}^{2}$ compensates for the gradient scaling introduced by temperature smoothing. The teacher is not an adversarial discriminator. It is frozen, multi-class, and provides training-only soft diagnostic supervision from the real follow-up radiograph. This avoids min--max instability while guiding the generator toward KL-relevant radiographic features. The distillation strength and temperature are selected empirically based on the fidelity--clinical consistency trade-off described in Sec.~\ref{sec:ablation}.

\subsection{Training Objective}
\label{sec:objective}

The final training objective combines pixel reconstruction, perceptual
similarity, and classifier distillation:
\begin{equation}
\label{eq:objective}
\begin{split}
    \mathcal{L}(\theta)
    =
    &\lambda_{\ell_2}
    \left\|
        \hat{I}_{v12} - I_{v12}
    \right\|_2^2
    +
    \lambda_{\mathrm{lpips}}
    \mathrm{LPIPS}
    \left(
        \hat{I}_{v12}, I_{v12}
    \right) \\
    &+
    \lambda_{\mathrm{distill}}
    \mathcal{L}_{\mathrm{distill}} .
\end{split}
\end{equation}

The reconstruction term encourages patient-specific structural preservation, the LPIPS term promotes perceptual similarity, and the distillation term aligns generated follow-ups with the diagnostic distribution of real follow-up radiographs. The loss weights are specified in the experimental setup. 
The final model is a non-adversarial SD2.1+LoRA radiographic forecasting model.
We disable adversarial training and do not include additional knee-VGG
feature-matching or latent-progression losses.

\section{Experiments and Analysis}
\label{sec:experiments}

\subsection{Experimental Setup}
\label{sec:experimental_setup}

\noindent\textbf{Dataset.}
We evaluate on the Osteoarthritis Initiative (OAI) longitudinal knee radiograph
cohort~\cite{nevitt2006osteoarthritis}. We retain knees with paired current
($v00$) and 12-month follow-up ($v12$) radiographs and available KL grades at
both visits, resulting in 5,798 knees from approximately 2,936 patients. We use
a patient-level split to avoid leakage: 3,835 knees from 1,942 patients for
training, 676 knees from approximately 338 patients for validation, and 1,287
knees from 656 patients for testing. CS-LDM receives only the current radiograph
$I_{v00}$ and current KL grade $y_{v00}$ as input; the future KL grade
$y_{v12}$ is used only for supervision and evaluation.

\noindent\textbf{Baselines.}
We compare CS-LDM with non-generative reference rows and reproduced generative
baselines. The V00 image floor evaluates the current radiograph against the
future label, while the Real V12 reference evaluates the true follow-up under
the same automated grading protocol. The Real V12 row is a reference for grader
behavior, not a strict upper bound on clinical validity. Generative baselines
include Pix2Pix~\cite{isola2017image}, CycleGAN~\cite{zhu2017unpaired},
Fast-DDPM~\cite{jiang2025fast}, GDM-VE~\cite{zhang2025geodesicdiffusionmodelsmedical},
IC-RDN~\cite{wu2024identity}, and HYPIR~\cite{lin2025harnessing}, which is
adapted as a restoration-style latent diffusion baseline without current KL
conditioning or diagnostic distillation. All methods are evaluated on the same held-out test split.

\noindent\textbf{Evaluation.}
We operationalize these two evaluation goals through four metric groups: image
fidelity, automated diagnostic consistency, anatomical-change fidelity, and
stable/progressed stress testing. Image fidelity is measured by PSNR, SSIM,
LPIPS, FID, and knee-FID. Diagnostic consistency is measured using
real-V12-trained KL graders across VGG-16, VGG-19, ResNet-18, ResNet-50, and
DenseNet-121, together with ordinal metrics including mAUC, QWK, and MAE.
Anatomical-change fidelity is evaluated using OARSI joint-space narrowing and
osteophyte agreement, zoomed ROI visualizations, and V00--V12 difference-map
analysis. Binary progression is defined as
\begin{equation}
    r^{i} = \mathbb{1}[y_{v12}^{i} > y_{v00}^{i}],
\end{equation}
and is treated as a secondary stress test due to the small number of true
12-month progressors.

\noindent\textbf{Implementation.}
All radiographs are cropped around the knee joint and resized to
$224 \times 224$. CS-LDM is trained for $30{,}000$ steps using Adam with learning
rate $2\times10^{-5}$, batch size 6, bfloat16 mixed precision, gradient
checkpointing, and gradient clipping. LoRA rank is set to 256. We use fixed
diffusion timestep $t=200$, $\lambda_{\ell_2}=1.0$,
$\lambda_{\mathrm{lpips}}=5.0$, $\lambda_{\mathrm{distill}}=0.3$, and
distillation temperature $\mathcal{T}=3.0$. The VGG-16 teacher is trained only
on real 12-month radiographs from the training split and then frozen.

\begin{table*}[t]
\centering
\caption{
Main comparison for longitudinal knee radiographic forecasting. Image-quality
metrics compare generated V12 radiographs with real V12 targets. Automated KL
consistency reports 5-class accuracy from independent real-V12-trained graders.
V00 image floor and Real V12 reference are non-generative rows and excluded from
ranking; Real V12 is a grader-behavior reference, not a strict clinical upper
bound. Best/second-best generative results are shown in \textbf{bold}/\underline{underline};
N/R denotes unavailable results.
}
\label{tab:main_comparison}
\resizebox{\textwidth}{!}{
\begin{tabular}{lccccc|ccccc}
\toprule
\multirow{2}{*}{Method}
& \multicolumn{5}{c|}{Image Quality}
& \multicolumn{5}{c}{Automated KL Consistency (\%)} \\
\cmidrule(lr){2-6}
\cmidrule(lr){7-11}
& PSNR $\uparrow$
& SSIM $\uparrow$
& LPIPS $\downarrow$
& FID $\downarrow$
& kFID $\downarrow$
& VGG16 $\uparrow$
& VGG19 $\uparrow$
& ResNet18 $\uparrow$
& ResNet50 $\uparrow$
& DenseNet121 $\uparrow$ \\
\midrule
V00 image floor
& 19.83 & 0.625 & 0.168 & -- & --
& \underline{61.54} & \underline{62.78} & \underline{55.40} & \underline{58.51} & 55.32 \\

Real V12 reference
& -- & 1.000 & 0.000 & 0.000 & 0.000
& 66.82 & 66.05 & 59.44 & 61.62 & 59.98 \\

Pix2Pix~\cite{isola2017image}
& \textbf{20.877} & \underline{0.700} & 0.191 & 38.966 & 25.182
& 47.71 & 42.11 & 36.60 & 39.01 & 33.88 \\

CycleGAN~\cite{zhu2017unpaired}
& 19.891 & 0.672 & 0.194 & \textbf{19.296} & 6.898
& \underline{61.54} & 57.65 & 45.92 & 50.19 & 45.69 \\

Fast-DDPM~\cite{jiang2025fast}
& 11.975 & 0.604 & 0.561 & 213.510 & 170.411
& 36.91 & 33.02 & 35.74 & 29.45 & 24.63 \\

GDM-VE~\cite{zhang2025geodesicdiffusionmodelsmedical}
& 19.856 & \textbf{0.746} & 0.328 & 110.387 & 19.912
& 58.43 & 60.14 & 39.94 & 55.94 & \underline{56.72} \\

IC-RDN~\cite{wu2024identity}
& 12.673 & 0.375 & 0.368 & 82.783 & 21.767
& 43.44 & 36.25 & 33.44 & 36.33 & 32.97 \\

\midrule
HYPIR (SD2.1 + LoRA)~\cite{lin2025harnessing}
& 20.680 & 0.698 & \textbf{0.135} & \underline{26.492} & \underline{4.300}
& 59.21 & 57.58 & \underline{55.40} & 55.79 & 56.41 \\

\textbf{CS-LDM(Ours)}
& \underline{20.704} & \underline{0.700} & \underline{0.139} & 29.701 & \textbf{1.644}
& \textbf{73.66} & \textbf{72.88} & \textbf{64.10} & \textbf{65.27} & \textbf{69.85} \\

\bottomrule
\end{tabular}
}
\end{table*}


\subsection{Diagnostic Consistency and Image Fidelity}
\label{sec:main_results}

Table~\ref{tab:main_comparison} compares CS-LDM with reference rows and
generative baselines. CS-LDM achieves the strongest automated KL consistency
across all five real-V12-trained graders, while maintaining competitive image
fidelity. Compared with HYPIR, CS-LDM improves VGG-16 KL accuracy from 59.21\%
to 73.66\%, and also obtains the best accuracy under VGG-19, ResNet-18,
ResNet-50, and DenseNet-121. This suggests that current KL conditioning and soft
diagnostic distillation improve future KL-relevant radiographic consistency
beyond generic latent restoration.

Image-quality metrics alone do not explain clinical usefulness. HYPIR obtains
the lowest LPIPS, and CycleGAN obtains the best generic FID, but their automated
KL consistency remains below CS-LDM. We therefore interpret CS-LDM's higher
automated KL accuracy as stronger diagnostic consistency, not as evidence that
generated images are clinically superior to real follow-up radiographs.
Full-image qualitative examples in the supplementary material further illustrate
this trade-off: some baselines over-smooth radiographs or preserve appearance
without producing KL-consistent follow-ups, whereas CS-LDM better balances
structural preservation and diagnostic consistency. The following sections
further evaluate whether the generated changes are anatomically meaningful.

Table~\ref{tab:ordinal_support} reports ordinal KL-grade agreement under the
fixed real-V12 VGG-16 grader. CS-LDM achieves the highest accuracy, mAUC, and
QWK, and the lowest MAE among generated methods. This suggests that the
generated follow-ups preserve KL severity ordering rather than matching discrete
KL labels. Since these metrics are computed using a fixed automated grader, we
interpret them as diagnostic consistency rather than evidence of clinical
superiority over real radiographs.

\begin{table}[t]
\centering
\caption{
Ordinal KL-grade agreement under the fixed real-V12 VGG-16 grader. These
teacher-derived metrics are reported as supporting evidence of automated
diagnostic consistency. 
}
\label{tab:ordinal_support}
\scriptsize
\setlength{\tabcolsep}{4pt}
\renewcommand{\arraystretch}{0.90}
\resizebox{\linewidth}{!}{
\begin{tabular}{lcccc}
\toprule
Method
& Acc. (\%) $\uparrow$
& mAUC (\%) $\uparrow$
& QWK (\%) $\uparrow$
& MAE $\downarrow$ \\
\midrule
\textit{V00 image}
& 61.54
& 85.75
& 76.65
& 0.4848 \\

\textit{Real V12}
& \underline{66.82}
& \underline{88.79}
& \underline{82.16}
& \underline{0.4025} \\

Pix2Pix~\cite{isola2017image}
& 47.71
& 78.65
& 59.47
& 0.7280 \\

CycleGAN~\cite{zhu2017unpaired}
& 61.54
& 85.94
& 77.60
& 0.4800 \\

Fast-DDPM~\cite{jiang2025fast}
& 36.91
& 66.16
& 32.80
& 0.9670 \\

GDM-VE~\cite{zhang2025geodesicdiffusionmodelsmedical}
& 58.43
& 83.42
& 72.60
& 0.5470 \\

IC-RDN~\cite{wu2024identity}
& 43.44
& 74.12
& 48.32
& 0.8320 \\

HYPIR~\cite{lin2025harnessing}
& 59.21
& 83.78
& 74.60
& 0.5220 \\

\textbf{CS-LDM}
& \textbf{73.66}
& \textbf{91.50}
& \textbf{87.30}
& \textbf{0.3030} \\
\bottomrule
\end{tabular}
}
\vspace{-0.4cm}
\end{table}

\subsection{Anatomical Feature Agreement}
\label{sec:oarsi_results}

Automated KL consistency measures whether generated follow-ups are interpretable
by real-image-trained graders, but it does not directly assess whether the
synthesized changes correspond to anatomical OA features. We therefore evaluate
feature-level agreement using an independent OARSI grading ensemble, which
estimates joint-space narrowing (JSN) and osteophyte-related sub-scores on both
generated and real follow-up radiographs. For each method, we compare the OARSI
sub-scores of the generated V12 image with those of the real V12 target.

Table~\ref{tab:oarsi} reports OARSI agreement on the full test set and
the progressed subset. HYPIR obtains the strongest JSN agreement and OARSI-based
KL accuracy, while CS-LDM achieves the best six-feature OARSI agreement on both
sets. This suggests that the restoration-style backbone preserves local
joint-space structure well, whereas CS-LDM provides stronger overall
feature-level alignment across JSN and osteophyte-related sub-scores. These
results provide anatomical evidence beyond classifier-based KL accuracy and
support the need to evaluate diagnostic consistency together with structural
feature agreement.

\begin{table}[t]
\vspace{-0.2cm}
\centering
\caption{
Independent anatomical evaluation using an OARSI grading ensemble. JSN and
6-mean report MAE between generated and real V12 radiographs; KL acc. is from
the same OARSI ensemble's KL head.
}

\label{tab:oarsi}
\scriptsize
\setlength{\tabcolsep}{2.5pt}
\renewcommand{\arraystretch}{0.90}
\resizebox{\linewidth}{!}{
\begin{tabular}{l|ccc|ccc}
\toprule
\multirow{2}{*}{Method}
& \multicolumn{3}{c|}{Full test set}
& \multicolumn{3}{c}{Progressed subset} \\
\cmidrule(lr){2-4}
\cmidrule(lr){5-7}
& JSN $\downarrow$
& 6-mean $\downarrow$
& KL acc. $\uparrow$
& JSN $\downarrow$
& 6-mean $\downarrow$
& KL acc. $\uparrow$ \\
\midrule
\textit{v00-copy}
& \textit{0.191}
& \textit{0.294}
& ---
& \textit{0.246}
& \textit{0.362}
& --- \\

CycleGAN~\cite{zhu2017unpaired}
& 0.477 & 0.459 & 0.716
& 0.554 & 0.508 & 0.615 \\

Pix2Pix~\cite{isola2017image}
& 0.402 & 0.442 & 0.848
& 0.469 & 0.505 & 0.892 \\

Fast-DDPM~\cite{jiang2025fast}
& 0.497 & 0.529 & 0.534
& 0.454 & 0.587 & 0.538 \\

GDM-VE~\cite{zhang2025geodesicdiffusionmodelsmedical}
& 0.415 & 0.484 & 0.724
& 0.462 & 0.556 & 0.769 \\

IC-RDN~\cite{wu2024identity}
& 0.339 & 0.490 & 0.815
& 0.365 & 0.556 & 0.857 \\

HYPIR~\cite{lin2025harnessing}
& \textbf{0.126} & \underline{0.266} & \textbf{0.948}
& \textbf{0.138} & \underline{0.297} & \textbf{0.938} \\

\textbf{CS-LDM}
& \underline{0.135} & \textbf{0.265} & \underline{0.934}
& \underline{0.154} & \textbf{0.269} & \underline{0.923} \\
\bottomrule
\end{tabular}
}
\vspace{-0.35cm}
\end{table}






\subsection{ROI-Level Change Localization}
\label{sec:roi_results}

Feature-level OARSI agreement summarizes anatomical consistency, but it does not
show where the predicted changes occur. We therefore evaluate ROI-level
longitudinal change on the progressed test knees, comparing the real change
$|I_{v12}-I_{v00}|$ with the predicted change
$|\hat{I}_{v12}-I_{v00}|$ in clinically relevant regions. Grad-$r$ measures
edge-pattern correlation, Edge-Dice measures local contour overlap,
$\Delta$-$r$ measures change-map alignment, ROI-LPIPS measures local perceptual
agreement with the real follow-up, and $\Delta$-L1 measures change-magnitude
error.

\begin{figure}[t]
    \centering
    \includegraphics[width=\linewidth]{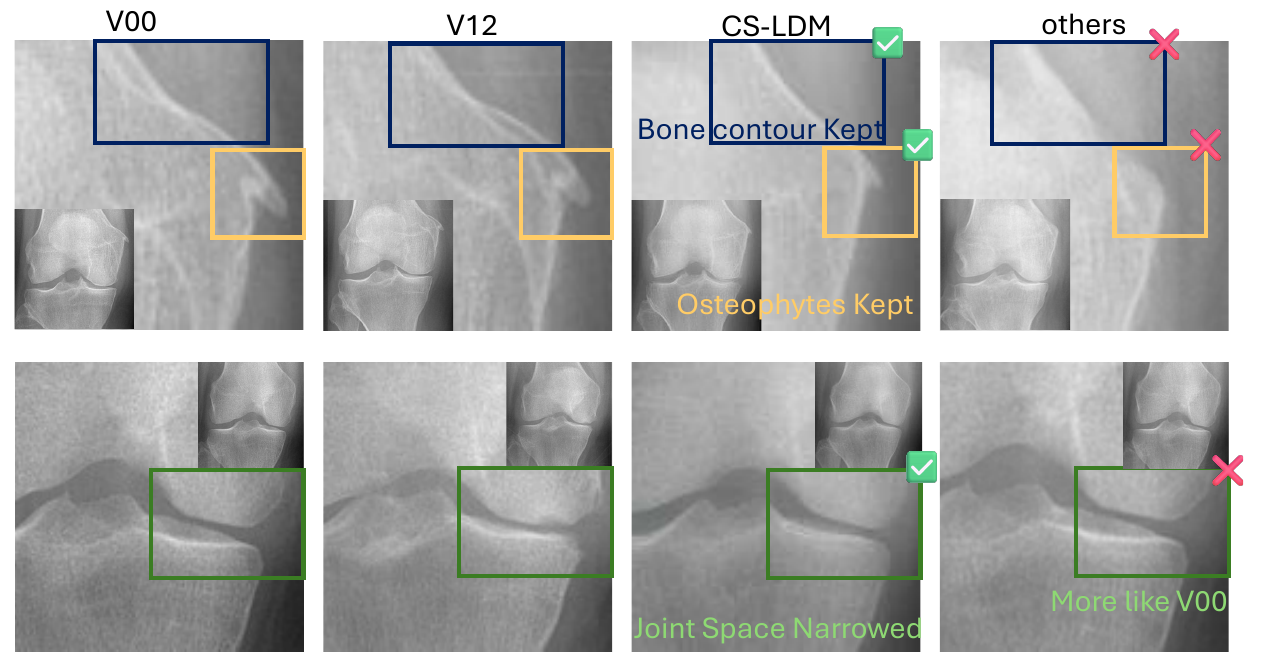}
    \caption{
    Representative ROI visualization on progressed cases. Boxes mark
    osteophyte-prone and joint-space regions; dataset-level metrics are in
    Table~\ref{tab:roi_metrics}.
    }
    \vspace{-0.5cm}
    \label{fig:roi_visualization}
\end{figure}

Figure~\ref{fig:roi_visualization} shows representative osteophyte-prone and
joint-space ROIs. The examples illustrate that CS-LDM can preserve local bone
morphology while producing localized V00-to-V12 changes. Since visual examples
can be case-dependent, we further report dataset-level ROI metrics on the common
progressed test cases in Table~\ref{tab:roi_metrics}.

Table~\ref{tab:roi_metrics} shows that CS-LDM is not uniformly best on every
local metric. HYPIR achieves the strongest contour agreement and ROI-LPIPS,
while Pix2Pix gives the best change-map correlation and lowest $\Delta$-L1.
CS-LDM remains competitive across both ROIs and obtains the second-best
ROI-LPIPS, outperforming all non-HYPIR baselines in local perceptual fidelity.
CycleGAN stays close to the V00-copy floor on several metrics, suggesting
limited anatomical change. Together with the OARSI results, the ROI analysis
supports a balanced view of CS-LDM: improved diagnostic and feature-level
agreement with competitive local morphology preservation.

\begin{table}[t]
\centering
\caption{
Dataset-level ROI metrics on progressed knees.
}

\label{tab:roi_metrics}
\tiny
\setlength{\tabcolsep}{1.8pt}
\renewcommand{\arraystretch}{0.78}
\resizebox{\linewidth}{!}{
\begin{tabular}{lccccc}
\toprule
Method
& Grad-$r$ $\uparrow$
& Edge $\uparrow$
& $\Delta$-$r$ $\uparrow$
& ROI-LPIPS $\downarrow$
& $\Delta$-L1 $\downarrow$ \\
\midrule
\multicolumn{6}{@{}l}{\textit{a: Lateral tibia osteophyte ROI}} \\
\textit{V00 copy}
& \textit{0.116} & \textit{0.191} & \textit{---} & \textit{0.140} & \textit{0.112} \\

Pix2Pix~\cite{isola2017image}
& 0.226 & 0.254 & \textbf{0.331} & 0.129 & \textbf{0.073} \\

CycleGAN~\cite{zhu2017unpaired}
& 0.117 & 0.181 & 0.016 & 0.150 & 0.093 \\

Fast-DDPM~\cite{jiang2025fast}
& 0.233 & 0.032 & -0.075 & 0.445 & 0.165 \\

GDM-VE~\cite{zhang2025geodesicdiffusionmodelsmedical}
& 0.243 & 0.232 & 0.168 & 0.332 & 0.089 \\

IC-RDN~\cite{wu2024identity}
& 0.202 & 0.217 & 0.172 & 0.415 & 0.205 \\

HYPIR~\cite{lin2025harnessing}
& \textbf{0.259} & \textbf{0.279} & \underline{0.289} & \textbf{0.106} & \underline{0.084} \\

\textbf{CS-LDM}
& \underline{0.247} & \underline{0.256} & 0.282 & \underline{0.114} & 0.085 \\
\midrule
\multicolumn{6}{@{}l}{\textit{b: Joint-space narrowing ROI}} \\
\textit{V00 copy}
& \textit{0.153} & \textit{0.203} & \textit{---} & \textit{0.155} & \textit{0.108} \\

Pix2Pix~\cite{isola2017image}
& 0.217 & 0.241 & \textbf{0.372} & 0.149 & \textbf{0.067} \\

CycleGAN~\cite{zhu2017unpaired}
& 0.157 & 0.193 & 0.063 & 0.164 & 0.090 \\

Fast-DDPM~\cite{jiang2025fast}
& 0.201 & 0.025 & -0.035 & 0.508 & 0.172 \\

GDM-VE~\cite{zhang2025geodesicdiffusionmodelsmedical}
& \underline{0.268} & 0.220 & 0.224 & 0.336 & \underline{0.083} \\

IC-RDN~\cite{wu2024identity}
& 0.187 & 0.213 & 0.167 & 0.440 & 0.195 \\

HYPIR~\cite{lin2025harnessing}
& \textbf{0.272} & \textbf{0.283} & \underline{0.292} & \textbf{0.128} & 0.086 \\

\textbf{CS-LDM}
& 0.244 & \underline{0.253} & 0.287 & \underline{0.134} & 0.086 \\
\bottomrule
\end{tabular}
}
\vspace{-0.35cm}
\end{table}

\subsection{Stable/Progressed Stress Test}
\label{sec:progression_stress}

Although CS-LDM improves automated KL consistency, 12-month KL progression is imbalanced: most test knees remain stable, while a small subset shows
a KL-grade increase. We therefore conduct a stable/progressed stress test using
an independent DenseNet-121 grader trained on real V12 radiographs. Progression
is defined as $y_{v12}>y_{v00}$. We also include a no-change V00-copy baseline,
which assumes the predicted V12 grade equals the current grade.

\begin{figure}[t]
    \centering
    \includegraphics[width=\linewidth]{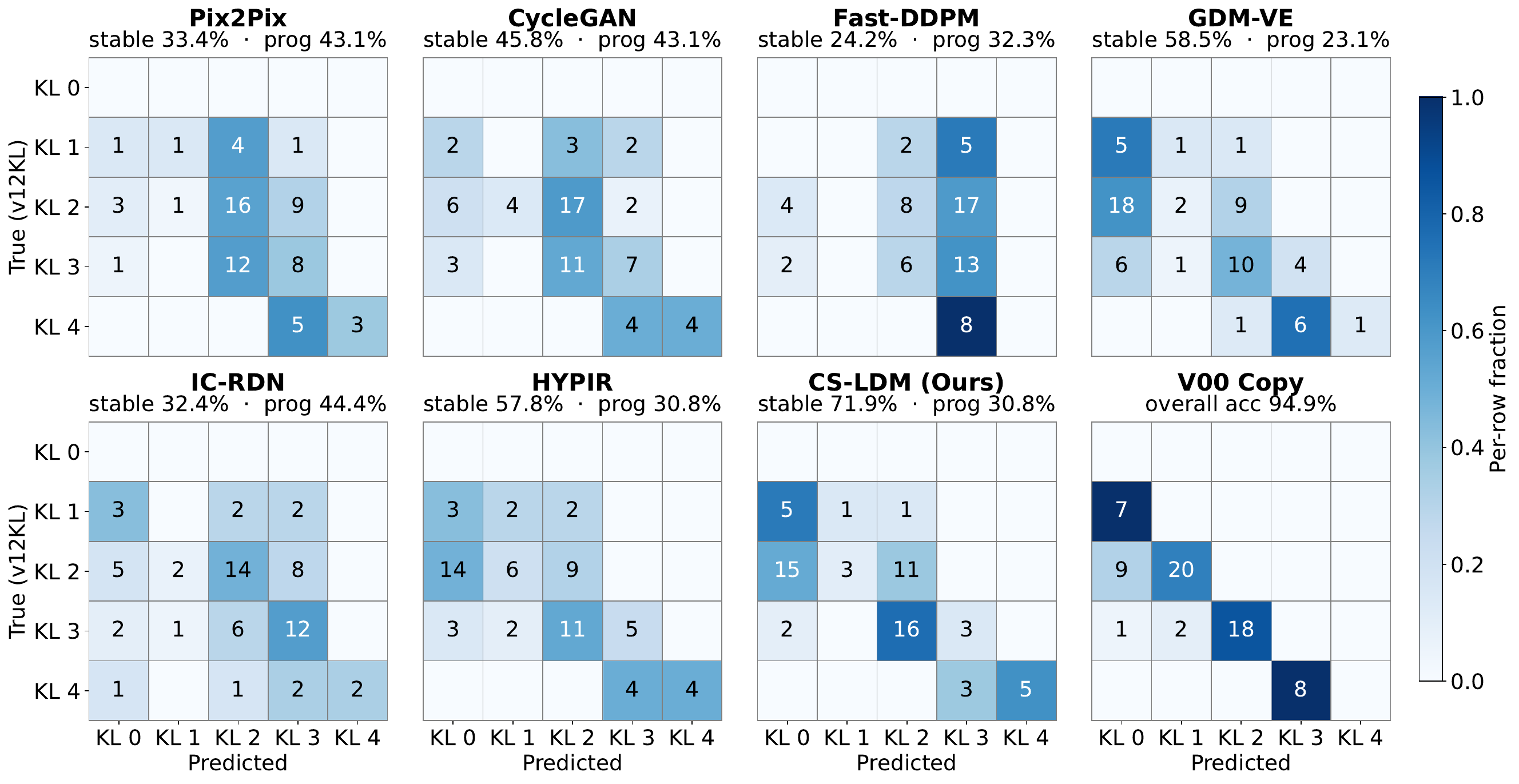}
    \caption{
    Progressed-subset confusion matrices under an independent DenseNet-121 grader.
    Rows are true V12 KL grades and columns are predicted KL grades. Stable/progressed
    accuracies are shown above each matrix. V00-copy assumes no V00-to-V12 KL change
    and fails on all true progressors by construction.
    }
    \vspace{-0.3cm}
    \label{fig:progression_cm}
\end{figure}

Figure~\ref{fig:progression_cm} shows progressed-subset confusion matrices for
the generated methods and the V00-copy baseline. The V00-copy baseline achieves
high overall accuracy because the test set is dominated by stable cases, but it
fails on every true progressed case by construction. Several baselines
show stronger progressed-case accuracy by collapsing predictions toward a small
number of KL grades. For example, Pix2Pix and CycleGAN frequently predict KL2,
while Fast-DDPM shifts many cases toward KL3. These behaviours improve selected
progressed counts but weaken stable-case preservation and ordinal severity
structure.

In contrast, CS-LDM achieves the highest stable-case accuracy (71.9\%) and the
best stable/progressed balanced accuracy (51.4\%) among the compared forecasting
methods, although its progressed-case accuracy remains modest (30.8\%). Balanced
accuracy is computed as the average of stable-case and progressed-case accuracy.
Its errors are concentrated within early OA grades, especially KL0--KL2, where
radiographic differences can be subtle and sensitive to acquisition angle or
positioning. Some KL2 cases are still under-called, indicating that rare
progression forecasting remains difficult. These results suggest that CS-LDM
better preserves stable cases and ordinal severity structure, but binary
progression forecasting over 12 months remains challenging under severe class
imbalance. We therefore treat binary progression detection as a secondary stress
test rather than the primary measure of forecasting quality.

\subsection{Ablation Study}
\label{sec:ablation}

We ablate three aspects of CS-LDM: current KL-grade prompting, frozen-teacher
diagnostic distillation, and the fixed timestep for single-step latent
forecasting. We further analyze teacher-architecture effects, while additional
distillation-strength and temperature sensitivity results are provided in the
supplementary material.

\noindent\textbf{Prompt and distillation synergy.}
Table~\ref{tab:prompt_distill_ablation} isolates the effects of current
KL-grade prompting and frozen-teacher distillation. Prompting alone gives only
a marginal improvement over the SD2.1+LoRA baseline, increasing accuracy from
0.5921 to 0.5952. Distillation alone improves accuracy to 0.6239, but degrades
LPIPS from 0.1354 to 0.1643, suggesting that diagnostic distillation can improve
diagnostic consistency at the cost of perceptual fidelity. Combining the current
KL prompt with light distillation achieves the highest accuracy of 0.7366 while
maintaining competitive LPIPS of 0.1386. This indicates that the final gain
comes from the interaction between clinically available current-grade
conditioning and soft diagnostic supervision.
\begin{table}[t]
\centering
\caption{Ablation of KL-grade prompting and distillation.}
\vspace{-0.25cm}
\label{tab:prompt_distill_ablation}
\scriptsize
\setlength{\tabcolsep}{5pt}
\renewcommand{\arraystretch}{0.86}
\resizebox{0.86\linewidth}{!}{
\begin{tabular}{lcccc}
\toprule
Method & Prompt & Distill & Acc. $\uparrow$ & LPIPS $\downarrow$ \\
\midrule
SD2.1+LoRA
& \xmark & \xmark
& 0.5921 & 0.1354 \\

+ Prompt only
& \cmark & \xmark
& 0.5952 & \textbf{0.1353} \\

+ Distill only
& \xmark & \cmark
& 0.6239 & 0.1643 \\

\textbf{CS-LDM}
& \cmark & \cmark
& \textbf{0.7366} & 0.1386 \\
\bottomrule
\end{tabular}
}
\end{table}

\noindent\textbf{Timestep sensitivity.}
Table~\ref{tab:t_ablation} evaluates the fixed diffusion timestep used for
single-step latent forecasting. The setting $t=200$ achieves the best accuracy,
QWK, MAE, and LPIPS among the tested timesteps. Lower timesteps provide weaker
diagnostic transformation, while higher timesteps can introduce stronger
image-level changes without improving ordinal consistency. We therefore use
$t=200$ as a structure-preserving forecasting timestep.

\begin{table}[t]
\centering
\vspace{-0.2cm}
\caption{Timestep sensitivity for single-step latent forecasting.}
\label{tab:t_ablation}
\scriptsize
\setlength{\tabcolsep}{2.5pt}
\renewcommand{\arraystretch}{0.88}
\resizebox{\linewidth}{!}{
\begin{tabular}{ccccc|ccccc}
\toprule
$t$ & Acc. $\uparrow$ & QWK $\uparrow$ & MAE $\downarrow$ & LPIPS $\downarrow$
&
$t$ & Acc. $\uparrow$ & QWK $\uparrow$ & MAE $\downarrow$ & LPIPS $\downarrow$ \\
\midrule
50
& 62.24 & 0.773 & 0.461 & 0.283
&
300
& 69.62 & 0.847 & 0.358 & 0.144 \\

100
& 71.48 & 0.859 & 0.331 & 0.183
&
500
& 65.03 & 0.793 & 0.441 & 0.164 \\

200
& \textbf{73.66} & \textbf{0.873} & \textbf{0.303} & \textbf{0.139}
&
-- & -- & -- & -- & -- \\
\bottomrule
\end{tabular}
}
\vspace{-0.5cm}
\end{table}

\noindent\textbf{Teacher-architecture trade-off.}
We further examine whether the distillation signal depends on the frozen teacher
architecture. Table~\ref{tab:cross_teacher_ablation} reports a cross-teacher
distillation matrix, where each row uses a different frozen teacher and each
accuracy column uses an independent real-V12-trained evaluator. The diagonal
entries are generally high, especially for ResNet-18 and ResNet-50 teachers,
indicating strong same-architecture alignment. However, these ResNet-based
teachers substantially degrade image fidelity, with worse LPIPS and kFID than
VGG-based teachers. In contrast, the VGG-16 teacher achieves the best LPIPS and
kFID while maintaining competitive accuracy across the other evaluators. We
therefore use VGG-16 as the default teacher, prioritizing a conservative balance
between diagnostic consistency and radiographic fidelity rather than maximizing
a single evaluator score.

\begin{table}[t]
\centering
\caption{
Cross-teacher distillation ablation. Diagonal accuracy entries use the same architecture for
distillation and evaluation, and should be interpreted carefully.
}
\label{tab:cross_teacher_ablation}
\scriptsize
\setlength{\tabcolsep}{2.2pt}
\renewcommand{\arraystretch}{0.88}
\resizebox{\linewidth}{!}{
\begin{tabular}{lccccc|ccc}
\toprule
\multirow{2}{*}{Teacher}
& \multicolumn{5}{c|}{KL Acc. (\%) $\uparrow$}
& \multicolumn{3}{c}{Quality} \\
\cmidrule(lr){2-6}
\cmidrule(lr){7-9}
& V16
& V19
& R18
& R50
& D121
& LPIPS $\downarrow$
& FID $\downarrow$
& kFID $\downarrow$ \\
\midrule
VGG-16
& \textbf{73.66}
& 72.88
& 64.10
& 65.27
& 69.85
& \textbf{0.1386}
& 29.70
& \textbf{1.64} \\

VGG-19
& 73.35
& \textbf{79.64}
& 68.30
& 69.31
& 68.07
& 0.1437
& \textbf{28.78}
& 2.47 \\

ResNet-18
& 75.76
& 78.79
& \textbf{88.42}
& 76.53
& 80.34
& 0.1575
& 30.40
& 10.35 \\

ResNet-50
& 76.07
& 79.56
& 75.68
& \textbf{85.86}
& 79.72
& 0.1688
& 31.61
& 19.15 \\

DenseNet-121
& 71.87
& 75.37
& 67.52
& 69.39
& \textbf{77.78}
& 0.1527
& 30.22
& 5.35 \\
\bottomrule
\end{tabular}
}
\vspace{-0.4cm}
\end{table}

\subsection{Clinical Interpretation and Limitations}
\label{sec:limitations}

CS-LDM outputs are synthesized radiographs and may contain subtle errors in
early-grade structures, particularly under acquisition-angle variation and
severe class imbalance. The stable/progressed analysis also shows that rare
12-month progression remains difficult to forecast reliably. Future work should
therefore include external validation, clinical reader studies, and
landmark- or registration-based measurements to assess longitudinal anatomical
change directly.

\section{Conclusion}

We presented CS-LDM, a clinically guided latent diffusion framework for
patient-specific 12-month knee radiographic forecasting. By conditioning on
clinically available current information and incorporating soft diagnostic
distillation, CS-LDM preserves patient-specific structure while synthesizing
future OA-related radiographic changes. Experiments on the OAI cohort show that
this design improves future KL-grade consistency while maintaining competitive
image fidelity, with anatomical analyses supporting the plausibility of the
predicted structural changes. These findings suggest that radiographic
forecasting should be assessed through both diagnostic consistency and
anatomical credibility, rather than image realism or grading performance alone.
Generated follow-up radiographs can therefore serve as interpretable predictive
representations that complement scalar prognosis, rather than substitutes for
real follow-up imaging or clinical diagnosis. Future work will further validate
this paradigm through external cohorts, reader studies, and direct measurements
of longitudinal structural change.








\bibliographystyle{IEEEtranN}
\bibliography{main}

\end{document}